\documentclass[onecolumn]{aa}
\usepackage{graphicx}
\usepackage{txfonts}
\usepackage{natbib}
\usepackage{deluxetable}
\usepackage{wasysym}
\begin{document}
   \title{Follow-up observations of binary ultra-cool dwarfs}

 \author{H. Bouy \inst{1,2}\thanks{Marie Curie Outgoing International Fellow (MOIF-CT-2005-8389)}, E.~L. Mart\'\i n \inst{2,3}, W. Brandner \inst{4}, T. Forveille\inst{5}, X. Delfosse\inst{5}, N. Hu\'elamo\inst{6}, G. Basri \inst{1}, J. Girard\inst{7}, M.-R. Zapatero Osorio\inst{2}, M. Stumpf \inst{4}, A. Ghez\inst{8}, L. Valdivielso\inst{2}, F. Marchis\inst{1}, A.~J. Burgasser \inst{9}, K. Cruz \inst{10}}

   \offprints{H. Bouy}

   \institute{Astronomy Dpt, 601 Campbell Hall, , UC Berkeley, Berkeley, CA 94720, U.S.A\\
              \email{hbouy@astro.berkeley.edu, basri@astro.berkeley.edu}
	      \and Instituto de Astrof\'\i sica de Canarias, c/ V\'\i a L\'actea S/N, E-38200 San Cristobal de La Laguna, Tenerife, Spain\\
              \email{ege@iac.es,lval@iac.es,mosorio@iac.es}
	      \and University of Central Florida, Department of Physics, PO Box 162385, Orlando, FL 32816-2385, U.S.A 
              \and Max-Planck Institut f\"ur Astronomie, K\"onigstuhl 17, D-69117 Heidelberg, Germany\\
              \email{brandner@mpia.de,stumpf@mpia.de}
	      \and Laboratoire d'Astrophysique, Observatoire de Grenoble, UJF, CNRS, BP 53, 38041 Grenoble, Cedex 9, France\\  
	      \email{Thierry.Forveille@obs.ujf-grenoble.fr,Xavier.Delfosse@obs.ujf-grenoble.fr}
	      \and Laboratorio de Astrof\'\i sca Espacial y F\'\i sica Fundamental (LAEFF-INTA), Apdo 50727, 28080 Madrid, Spain\\
	      \email{nhuelamo@laeff.inta.es}
              \and Universidad Nacional Autn\'onoma de M\'exico, Apartado Postal 72-3, 58089 Morelia, Michoac\'an, Mexico\\
              \email{girard@astroscu.unam.mx}
	      \and Department of Physics and Astronomy, UCLA, Los Angeles, CA-90095-1547, U.S.A\\
             \email{ghez@astro.ucla.edu}
	      \and MIT Kavli Institute for Astrophysics \& Space Research, 77 Massachusetts Ave., Building 37-664B, Cambridge, MA 02139, MA-02139, U.S.A.  \\
	      \email{ajb@mit.edu}
	      \and Californian Institute of Technology, MC 105-24. 1200 East California Blvd, Pasadena CA-91125\\
	      \email{kelle@astro.caltech.edu}
             }

   \date{Received ..., 2007; accepted , 2007}

  \abstract
  % context heading (optional)
  % {} leave it empty if necessary
   {Astrometric observations of resolved binaries provide estimates of orbital periods and will eventually lead to measurement of dynamical masses. Only a few very low mass star and brown dwarf masses have been measured to date, and the mass-luminosity relation still needs to be calibrated.}
  % aims heading (mandatory)
   {We have monitored 14 very low mass multiple systems for several years to confirm their multiplicity and, for those with a short period, derive accurate orbital parameters and dynamical mass estimates.}
  % methods heading (mandatory)
   {We have used high spatial resolution images obtained at the Paranal, Lick and HST observatories to obtain astrometric and photometric measurements of the multiple systems at several epochs. The targets have periods ranging from 5 to 200 years, and spectral types in the  range M7.5 -- T5.5. }
  % results heading (mandatory)
   {All of our 14 multiple systems are confirmed as common proper motion pairs. 
One system  (2MASSW~J0920122+351742) is not resolved in our new images, probably 
because the discovery images were taken near maximum elongation. Six systems have periods 
short enough to allow dynamical mass measurements within the next 15 to 20~years. 
We estimate that only 8\% of the ultracool dwarfs in the solar neighborhood 
are binaries with separations large enough to be resolved, and yet periods short enough to derive astrometric orbital fits over a reasonable time frame with current instrumentation. A survey that doubles the number of ultracool dwarfs 
observed with high angular resolution is called for to 
discover enough binaries for a first attempt to derive the mass-luminosity 
relationship for very low-mass stars and brown dwarfs.}
  % conclusions heading (optional), leave it empty if necessary
   {}


\authorrunning{Bouy et al.}
\titlerunning{Follow-up observations of binary ultra-cool dwarfs}

   \maketitle

\section{Introduction}

Over the last few years, intensive computational and observational efforts have been made to improve our understanding of the formation processes and evolution of brown dwarfs 
(BDs) and very low mass (VLM) stars. The determination of their Initial Mass Function (IMF) is a crucial step in this direction. Translating an observed luminosity function into an IMF requires an accurate determination of their mass-luminosity relationship at different ages, which up to now relies primarily on theoretical mass-luminosity  relationships. Although the empirical constraints on these relationships for VLM stars have considerably improved within the past years \citep[see e.g][]{2004ApJ...604..741H,2000A&A...364..217D,2000A&A...364..665S} only a few observational constraints are currently available and large uncertainties remain \citep[][]{2001A&A...367..183L, 2004A&A...423..341B, 2004A&A...428..205B, 2004ApJ...615..958Z,2005Natur.433..286C,2006Natur.440..311S,2008arXiv0801.1525I}. 

The degeneracy in the mass-luminosity relation for ultra-cool dwarfs (UCDs) makes it difficult to accurately estimate their physical properties. Dynamical masses, which are not model-dependent, are a unique way to calibrate this relation.  The components of a multiple system are expected to be coeval, removing part of the above mentioned degeneracy. Although the ages of the targets studied in this work are not well constrained, it will be possible, once their dynamical masses are known, to take advantage of their coevality to test the evolutionary models. By adjusting the theoretical isochrones empirically to fit both the observed total masses and the individual luminosities of the multiple systems, it will be possible to directly check the consistency of the models with the observations. The corresponding predictions on the age can then be compared to other indicators such as the activity, the rotation, and the presence and strength of particular spectral features (such as Li, H$\alpha$), but also to more recent techniques based on spectral analysis of gravity sensitive features as described by \citet{2004ApJ...609..885M}, \citet{2004ApJ...600.1020M}, \citet{2003ApJ...593L.113M}, \citet{2004ApJ...615..958Z} and \citet{2006ApJ...639.1095B}. Finally, by studying the physical characteristics of objects with known dynamical masses, it will be possible to provide crucial information for our understanding of their physical properties, such as their interior structure, the formation of dust, the settling and depletion of refractory elements, and the underlying opacities. An accurate determination of the mass of an object based on dynamical masses in binary systems therefore provides not only a reality check for the theory but also a cornerstone in the understanding of the mass distribution of brown dwarfs.

In this work, we present a time-series of high angular resolution observations aimed at monitoring binary ultra-cool dwarfs. These observations confirm the common proper motion of the binary candidates and represent a first step towards the derivation of orbital parameters and dynamical masses. Most objects presented here were monitored over timescales too short in comparison with their periods, allowing us to estimate 
rough orbital periods, but preventing us from obtaining detailed orbital fits. 

\section{Observations and data analysis\label{observations}}
We have monitored 14 multiple systems using a variety of instruments; in the Hubble Space Telescope (HST), we used the Wide Field Planetary Camera (WFPC2), the Space Telescope Imaging Spectrograph (STIS), the Near-Infrared Camera and Multi-Object Spectrometer (NICMOS) and the Advanced Camera for Surveys (ACS). From the ground, we collected observations using the Very Large Telescope (VLT) with the NACO adaptive optics system, and the Lick Observatory Shane 3-m telescope, also with an adaptive optics system. Table \ref{instruments} gives an overview of the characteristics of these instruments. The observations reported here started in 2000, and continued until the end of 2007. Table \ref{results} lists the observations recorded per target, together with the corresponding relative astrometry, and photometry analysis.

\subsection{Sample}
The sample includes 14 binaries ranging from spectral M7.5 to T5.5 (see Table \ref{sample}), therefore covering a wide range of primary masses. All objects were known to be multiple and had been observed and resolved at least once \citep[see][]{2001AJ....121..489R, 2003AJ....126.1526B,2003AJ....125.3302G,2003ApJ...587..407C,2003ApJ...586..512B,2005ApJ...621.1023S}.

\subsection{HST/WFPC2 observations}
We used the HST/WFPC2 and its Planetary Camera \citep[PC, ][]{WFPC..Instrument..Handbook} with the F814W filter (programs GO-9157,GO-9345,GO-9499 and GO-9968, P.I. E.~L. Mart\'\i n). Part of these datasets were published in \citet{2004A&A...428..205B} and \citet{2005AJ....129..511B}. All targets were centered in the Planetary Camera (PC) which provides the best sampling of the PSF. In order to identify and remove cosmic ray events and bad pixels, we used a four-point dithering pattern with typical exposure times of 400~s, adding up to a total of 1600~s in total for each target. The images have been processed following standard procedures using the STScI STSDAS package in IRAF together with the calibration files provided by the STScI team.

\subsection{HST/NICMOS observations}
We retrieved NICMOS data from the HST public archive (program 9843, P.I. Gizis). Two objects (2MASSW~J0850359+105715 and 2MASSW~J1728114+394859) have been observed using the NICMOS1 camera. The objects were observed in MULTIACCUM mode with exposure times of 128~s and 144~s respectively. The data were processed following the recommendations of the HST Data Handbook with the STSDAS pipeline in IRAF and standard STScI calibration files.

\subsection{HST/ACS observations}
We started using the HST/ACS and its High Resolution Channel camera \citep[HRC,][]{ACS_INSTR_HANDBOOK} with the F625W, F775W and F850LP filters (program GO-9451, P.I. Brandner). Later observations were collected using only the F814W filter (GO-10559. P.I. Bouy) to obtain additional epochs for known VLM binaries. Part of the earlier datasets have already been published in \citet{ 2004A&A...424..213B,2004A&A...423..341B}. The data were obtained in CR-SPLIT mode with a four points dithering pattern in each filter, and typical exposure times of 490~s, 230~s, 180~s and 300~s with the F625W, F775W, F850LP and F814W filters respectively. The images have been processed following standard procedures using the STScI STSDAS package in IRAF together with the calibration files provided by the STScI team. 

\subsection{HST/STIS observations}
As part of program GO-9451 (P.I. Brandner), spatially resolved STIS spectra of binaries of this sample were obtained using the high spatial resolution STIS spectrograph on-board HST \citep{STIS_INSTR_HANDBOOK}. The corresponding spectroscopic data have been described in detail in another paper \citep{2006A&A...456..253M}. This paper focuses only on the pre-acquisition images obtained with STIS prior to each spectroscopic exposure. These images were obtained in the Longpass filter ($\lambda_{\rm cen}=$7230\AA, FWHM=2720\AA), with typical exposure times of 5 to 10~s. They have been processed following standard procedures as described in the STIS User's Handbook using the STScI STSDAS package in IRAF together with the calibration files provided by the STScI team.

\subsection{VLT/NACO observations}
We used the adaptive optics system NACO in order to obtain high spatial resolution images of VLM binaries (programs 70.D-0773, 077.C-0062, 71.C-0327, P.I. Bouy). NACO and its near-infrared wavefront sensor provided excellent diffraction limited images of the binaries. Prior to period 71, we requested to use the AO system with the N20C80 dichroic. This dichroic allows 80\% of the near-infrared light to reach the NIR wavefront sensor and 20\% to be collected by the ALADDIN detector of the science camera. After period 71, all images have been obtained with the N90C10 dichroic that sends 90\% of the light to the adaptive optics and 10\% to the ALADDIN detector, allowing to close the loop on even fainter objects. Our scientific targets were used as reference star for the wavefront sensing. The images were obtained in jitter mode with a four or five points dithering pattern. We processed the data with the recommended Eclipse \emph{jitter} package \citep{1997Msngr..87} and the calibration files provided by the Paranal observatory.

\subsection{Lick/AO observations}
We used the adaptive optics facility of the Lick Observatory Shane 3~m telescope  \citep{2002SPIE.4494..336G} on 2007 April 4$^{\rm th}$ to observe 2MASS~J1847034+552243 (using H and Ks broad-band filters) and 2MASS~J1047138+402649 (Ks only). These two targets and their neighboring stars are too faint to be used as natural guide star for Lick/AO wavefront sensing. Thus, we used the AO together with the Laser Guide Star (LGS) system. The Lick LGS system can perform tip-tilt wavefront sensing on a reference star brighter than R$<$16.5~mag. For the tip-tilt wavefront sensing, we used USNO-B1.0~1304-0211669 ($\alpha$=10h47min12.61s, $\delta$=+40\degr26min44.0s, R=16.5~mag) and USNO-B1.0~1453-0276611 ($\alpha$=18h47min00.6s, $\delta$=+55h22min25.3s, R=15.6~mag), located at 14\farcs4 and 29\farcs8 of 2MASS~J1047138+402649 and 2MASS~J1847034+552243, respectively. The laser spot was used for higher order corrections. The targets were observed using a 5 point dithering pattern, with exposure times of 30~s at each position. A PSF reference star was obtained just after 2MASS~J1847034+552243. In the case of 2MASS~J1047138+402649, we used the first component of the system as reference PSF, ensuring optimized results 
for the PSF fitting procedure.

\subsection{Analysis of the data \label{psf_fitting}}
In order to obtain the precise relative astrometry of these multiple systems, we used the same software described by \citet{2003AJ....126.1526B}, adapted to ACS, STIS, NICMOS, Lick/AO and VLT/NACO. The program, its performances and limitations are fully described in the paper cited above. 
A single point source can be described by only three parameters: the position of its centroid ($x,y$), and its total flux ($f$). A binary system is described by 6 parameters. The custom made program makes a non-linear fit of the binary system, fitting both components simultaneously rather than individually. It uses a library of 10 reference PSF \citep[9 natural PSF and 1 TinyTIM synthetic PSF in the case of HST, ][]{TINY_TIM}, except in the case of Lick/AO and NACO, for which only one reference PSF star obtained the same night with the same instrumental settings was used. A $\chi^{2}$--minimization between the synthetic binary and the observed binary gives the best values for the six parameters. Typical uncertainties and systematic errors are described in \citet{2003AJ....126.1526B} and \citet{2004PhDT.........5B} for both ACS and WFPC2. Similar calibrations have been done for Lick/AO and NACO. Briefly, for well resolved multiple systems with moderate differences of magnitude, uncertainties and systematic errors are in general estimated to add up to $\approx$10\% of the plate-scale of the instrument, provided that 3 conditions are met: a) the PSF is well sampled, b) the reference PSF is of good quality and c) that the signal-to-noise ratio is large enough. The effective resolution also depends on the technique used to measure the relative astrometry and photometry \citep[see e.g][ for 3 independent techniques]{2002ApJ...567L..53C,2003AJ....126.1526B,2005ApJ...633..452K}. The values quoted in Table \ref{instruments} are only indicative and relatively conservative. These uncertainties do not include systematic instrumental errors, which are discussed below and can sometimes dominate. Table \ref{systematics} gives an overview of these systematic errors. They should be added quadratically to the uncertainties given in Table \ref{results}. 

\subsubsection{HST/WFPC2 systematic errors}
The main systematic errors on relative astrometry are due to:
\begin{itemize}
\item the uncertainty on the absolute roll angle of the spacecraft ($<$0.003\degr\, according to the User's manual)
\item 34$^{\rm th}$ row defect producing an astrometric offset of approximately 3\% of the pixel height every 34 rows
\item the geometric distorsion ($<$0\farcs005 of error according to the User's manual)
\end{itemize}
The separations of the multiple systems presented in this paper are all less than 13 rows, so that the 34$^{\rm th}$ row defect affects them once at most. The maximum systematic errors on the relative astrometry measured with WFPC2 therefore adds up to 0\farcs0052, and the position angle to 0.003\degr.

\subsubsection{HST/ACS systematic errors}
The systematic errors are primarily due to the accuracy of the roll angle of the spacecraft ($<$0.003\degr\, as above) and to the accuracy with which the geometric distortion of the camera has been characterized. The MultiDrizzle \citep{MultiDrizzle} pipeline corrects for most of the geometric distorsions, and the final relative astrometry is expected to be better than 0.1 pixel, or $\approx$0\farcs0028.

\subsubsection{HST/STIS systematic errors}
As in the case of WFPC2 and ACS, the systematic errors are mainly due to the accuracy of the orientation of the spacecraft ($<$0.003\degr\, as above) and to the stability of optical distortion. The STIS Instrument Handbook gives an accuracy for relative astrometry within an image better than 0.1~pixel, corresponding to $\approx$5.1~mas \citep{STIS_INSTR_HANDBOOK}.

\subsubsection{HST/NICMOS1 systematic errors}
The NICMOS pixel scales along the X and Y axes of each camera are slightly different, because of the slight tilt of the NICMOS arrays relative to the focal plane. The difference is of the order of 3\permil\, only, and we neglect it in our analysis. The distortion corrections for the NICMOS1 camera are small, even at the edge of the camera (0.9 pixels). After correction using the \emph{drizzle} package provided by the STSci team, the relative astrometry in the center of the camera where all our targets were observed is expected to be better than 0.1~pixel corresponding to 4.4~mas. As in the case of the other HST instruments, the systematic errors also include the uncertainty on the orientation of the spacecraft \citep[$<$0.003\degr\, as above][]{NICMOS..Instrument..Handbook}. 
 
\subsubsection{VLT/NACO and Lick/AO  systematic errors}
In addition to static instrumental uncertainties, images obtained with AO are known to suffer from variable effects, due in particular to temporal and spatial variability of the atmospheric conditions. These effects can vary significantly on short timescales, 
even between two consecutive exposures, and thus a recorded PSF is only an approximation of the system's PSF. In the case of Lick/AO, we measured the effective platescale and position angle using a set of astrometric calibrators. The platescale was found to vary by as much as 1\%, corresponding to 0.8~mas/pixel, and the position angle to be off by as much as 0.34\degr. Our NACO observations were made in service mode with standard calibrations, and no astrometric calibrators were therefore obtained to control the platescale and orientation accuracy. \citet{2007A&A...474..273E} report recent measurements of the instrumental uncertainties obtained for NACO with similar settings. They measure platescale variations as large as 1\%, corresponding to 0.14~mas/pixel, and position angle offsets as large as 0.31\degr.

\begin{center}
\begin{deluxetable}{lcc}
\tabletypesize{\small}
\tablecaption{Estimates of the maximum systematic astrometric errors obtained with HST, VLT/NACO and Lick/AO \label{systematics}}
\tablewidth{0pt}
\tablehead{
\colhead{Instrument} & \colhead{Error Sep.} & \colhead{Error P.A} \\
\colhead{}           & \colhead{}      & \colhead{[\degr]} 
}
\startdata
HST/WFPC2            & 5.2~mas          & 0.003 \\
HST/ACS              & 2.8~mas          & 0.003 \\
HST/STIS             & 5.1~mas    & 0.003 \\
HST/NICMOS           & 4.4~mas    & 0.003 \\
VLT/NACO             & 1\%              & 0.31   \\
Lick/AO              & 1\%              & 0.35  \\
\enddata
\end{deluxetable}
\end{center}

\section{Analysis} 

\subsection{Common proper motion pairs}
Ten objects have proper motion measurements in \citet{2007arXiv0710.4786J}, \citet{2002AJ....124.1170D}, \citet{2003AJ....126..975T}, \citet{2004AJ....127.2948V} or the USNO-B.1 catalog \citep{2003AJ....125..984M}. All but three of these targets are confirmed as common proper motion pairs with motion of the secondary much lower than the proper motion (see Table \ref{motion}). For the L-dwarf pairs 2MASSW~2331016-040619 and 2MASSW~J1728114+394859, Table \ref{motion} gives a proper motion amplitude comparable to the motion of the secondary, but the orientation of the proper motion of the unresolved pairs is inconsistent with the companion being an unrelated background source, as illustrated in Fig. \ref{orbits}. The L dwarf 2MASSW~J0920122+351742 is not resolved in our new VLT and HST images (see Section \ref{2m0920} for a detailed discussion on that particular object). Even though accurate kinematics measurements are required to confirm that the objects without proper motion measurements are comoving, we note that the motion of the secondary component with respect to the primary is consistent with that expected for a gravitationally-bound companion. Considering the uncertainties, the relative motion is of the order of $\approx$10~mas/yr, typically lower than the proper motions expected for such nearby objects \citep[$\approx$100~mas/yr, see Table \ref{motion} and e.g][]{2002AJ....124.1170D,2003AJ....126..975T}.

\subsection{2MASSW~J0920122+351742 \label{2m0920}}

2MASSW~J0920122+351742 (L6.5) has been unambiguously resolved as a binary by \citet{2001AJ....121..489R} using HST/WFPC2, with a separation of 0\farcs075. This object is not resolved by us, neither in our 2 epochs with HST/ACS and HST/STIS, nor in our third VLT/NACO epoch. Figure \ref{2m0920} shows a mosaic of the 6 epoch images of 2MASSW~J0920122+351742 obtained with HST and VLT. 

\begin{figure*}
\begin{center}
\includegraphics*[width=0.6\textwidth]{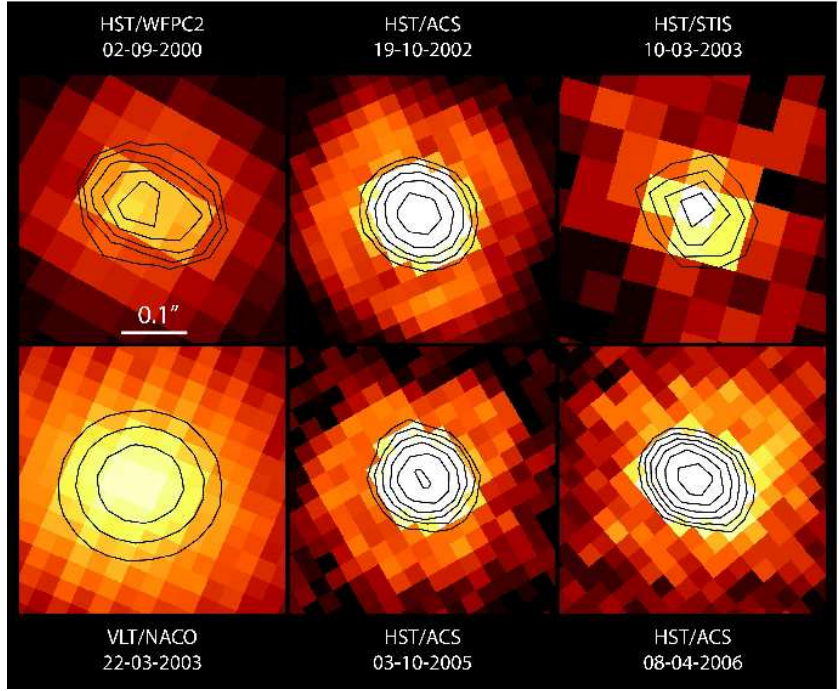}
\caption{Mosaic of images of 2MASSW~J0920122+351742. The observation date and instrument are indicated. The scale is represented and is the same in each image stamp. Contour plots are over-plotted to illustrate the clear elongation in the first epoch image \citep{2001AJ....121..489R}, the possible elongation in the last epoch image in the same direction, and the round PSF at the other epochs.  \label{contour_2m0920}}
\end{center}
\end{figure*}

The object is clearly elongated in the WFPC2 image, as shown in Figure \ref{contour_2m0920}. It is elongated in the three consecutive images obtained that day, excluding the possibility of a cosmic ray event. Moreover, other objects present in the field of view of the WFPC2 images do not show any elongation, excluding any instrumental problem. 

The presence of the nearby star 2MASS~J09201092+3517452 in the February 2000 WFPC2 image and in the March 1998 2MASS images allows us to rule out the combination of a high proper motion brown dwarf with a background star aligned by chance at the first epoch.

Figure \ref{contour_2m0920} shows that the PSF of the 4 consecutive ACS (2002), STIS (2003), NACO (2003) and ACS (2005) images look sharp and unresolved.  Because the system is not resolved, we can put an upper limit of $\approx$0\farcs06 on the separation of the two components of the system, corresponding over the 5.6~yr time difference to a motion of 0\farcs011~yr$^{-1}$. This measured motion is  much smaller than the typical 0\farcs100~yr$^{-1}$ reported for such nearby ultracool dwarfs \citep[see Table \ref{motion}, and ][]{2002AJ....124.1170D,2003AJ....126..975T}, and suggests that the absence of motion detection is due to the fact that the pair is comoving (assuming negligible motion for eventual background coincidence). An accurate proper motion measurement should confirm this preliminary conclusion.

\begin{figure*}
\begin{center}
\includegraphics*[width=0.6\textwidth]{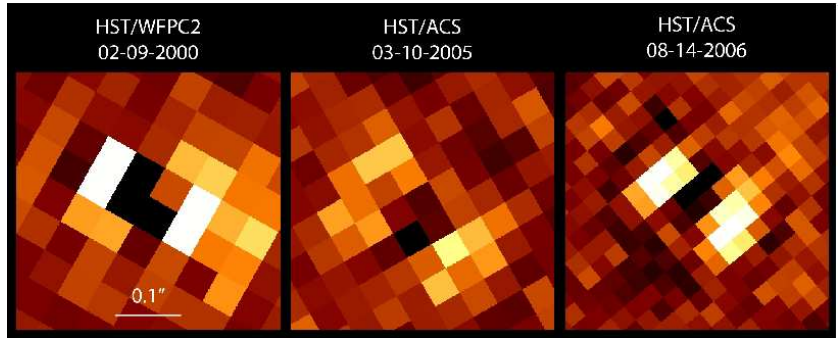}
\caption{Comparison of the average residuals obtained after single-star PSF subtraction on the resolved WFPC2 image of 2MASSW~J0920122+351742 (left), the unresolved ACS image (middle), and the possibly resolved ACS image (right). The color scale and orientation (North/Up and East/Left) are the same in each image. The scale is indicated in the left stamp and is the same for each image. The residuals are significantly stronger in the 2000 WFPC2 and in the 2006 ACS image than in the 2005 ACS image.   \label{residuals_2m0920}}
\end{center}
\end{figure*}

A more detailed analysis of the last epoch image (2006) obtained with ACS shows that the PSF seems a little elongated. Figure \ref{residuals_2m0920} shows a comparison of the residuals after PSF subtraction of the resolved WFPC2 image, the unresolved ACS image of 2005 and the possibly resolved last epoch ACS image. The residuals are significantly stronger in the first and last one, with an elongation in the same direction, indicating that the object is possibly almost resolved in the last epoch. The first and last epochs are separated by 5.6~yr, close to the estimated orbital period \citep[$\approx$7.2 years,][]{2003AJ....126.1526B}. This suggests as possible explanation that the companion might have been too close to be resolved in the NACO, ACS (2003 and 2005) and STIS images, while close to its maximum elongation in the WFPC2 and last ACS image. The relatively short estimated period of $\approx$7.2 years, and the short separation (only 0\farcs075, very close to the limit of resolution of HST and VLT at these wavelengths) are consistent with such a scenario. 

\begin{figure*}
\begin{center}
\includegraphics*[width=\textwidth]{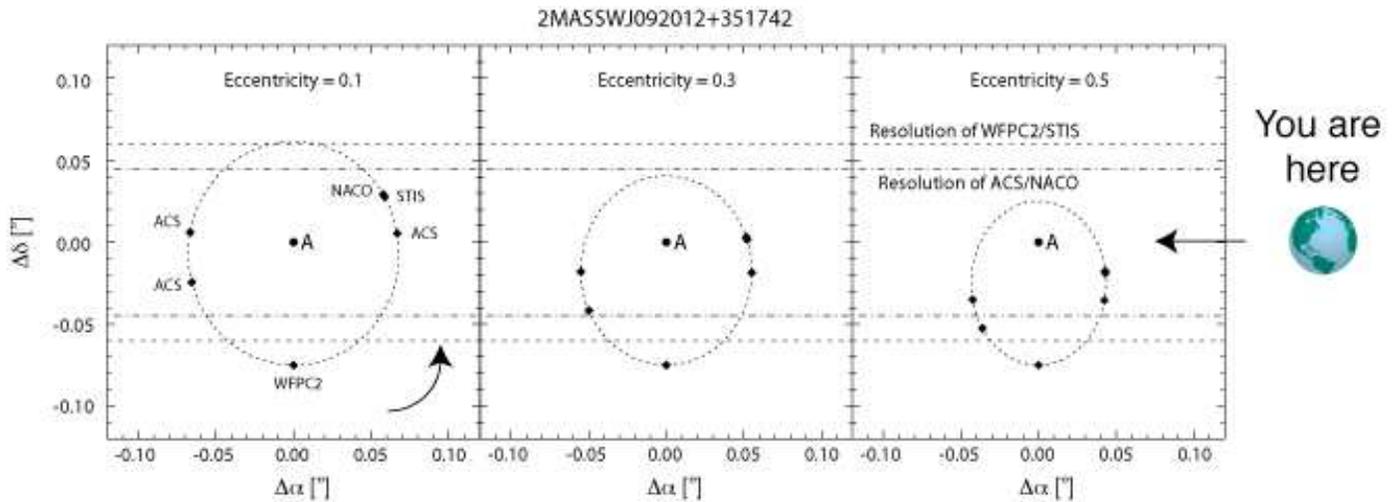}
\caption{Figure illustrating possible scenarios explaining why the companion was detected in the HST/WFPC2 images but not in the VLT/NACO, HST/STIS and HST/ACS ones. It assumes orbits with typical eccentricities of 0.1, 0.3 and 0.5 for 2MASSW~J0920122+351742AB, with a semi-major axis of 0\farcs075, a period of 7.2~yrs, and the apastron at the epoch of the WFPC2 observations. The filled circle in the center indicates the position of the primary, and the diamonds the position of the secondary at the epochs of the 5 observations. The horizontal dashed and dashed-dotted lines indicate the limit of resolution of respectively WFPC2/STIS ($\approx$0\farcs060), and ACS/NACO ($\approx$0\farcs045), as estimated in \citet{2004PhDT.........5B,2003AJ....126.1526B}. The line of sight, chosen in the most favorable case, is indicated on the right.\label{2m0920sim}}
\end{center}
\end{figure*}

Simple calculations considering an eccentric orbit, with a period of 7.2~yrs, a semi-major axis of 0\farcs075, as measured in the WPFC2 image, and the companion at its apastron at the date the WFPC2 images indicate that the probability that the companion could not be resolved by either NACO, STIS or ACS is relatively high. Figure \ref{2m0920sim} illustrates these calculations in the cases of typical eccentricities  of 0.1, 0.3 and 0.5. In these configurations, and for eccentricities greater than 0.3, the companion would have been resolved (or almost resolved) in the last ACS image but in none of the other ACS, STIS or NACO images. Although simplistic, these calculations show that further observations of 2MASSW~J0920122+351742 will have to be taken near maximum elongation in order to resolve the binary again with currently available instruments.

%Although the probability is extremely low, 2MASSW~J0920122+351742 could be an eclipsing binary. We used the 3 epoch HST images obtained in the F814W filter to search for variability. The luminosity of 2MASSW~J0920122+351742 does not show any significant variations. A careful photometric monitoring is nevertheless required to rule out this possibility.

\subsection{Discussion on particular objects}

\begin{figure*}
\begin{center}
\includegraphics*[height=0.9\textheight]{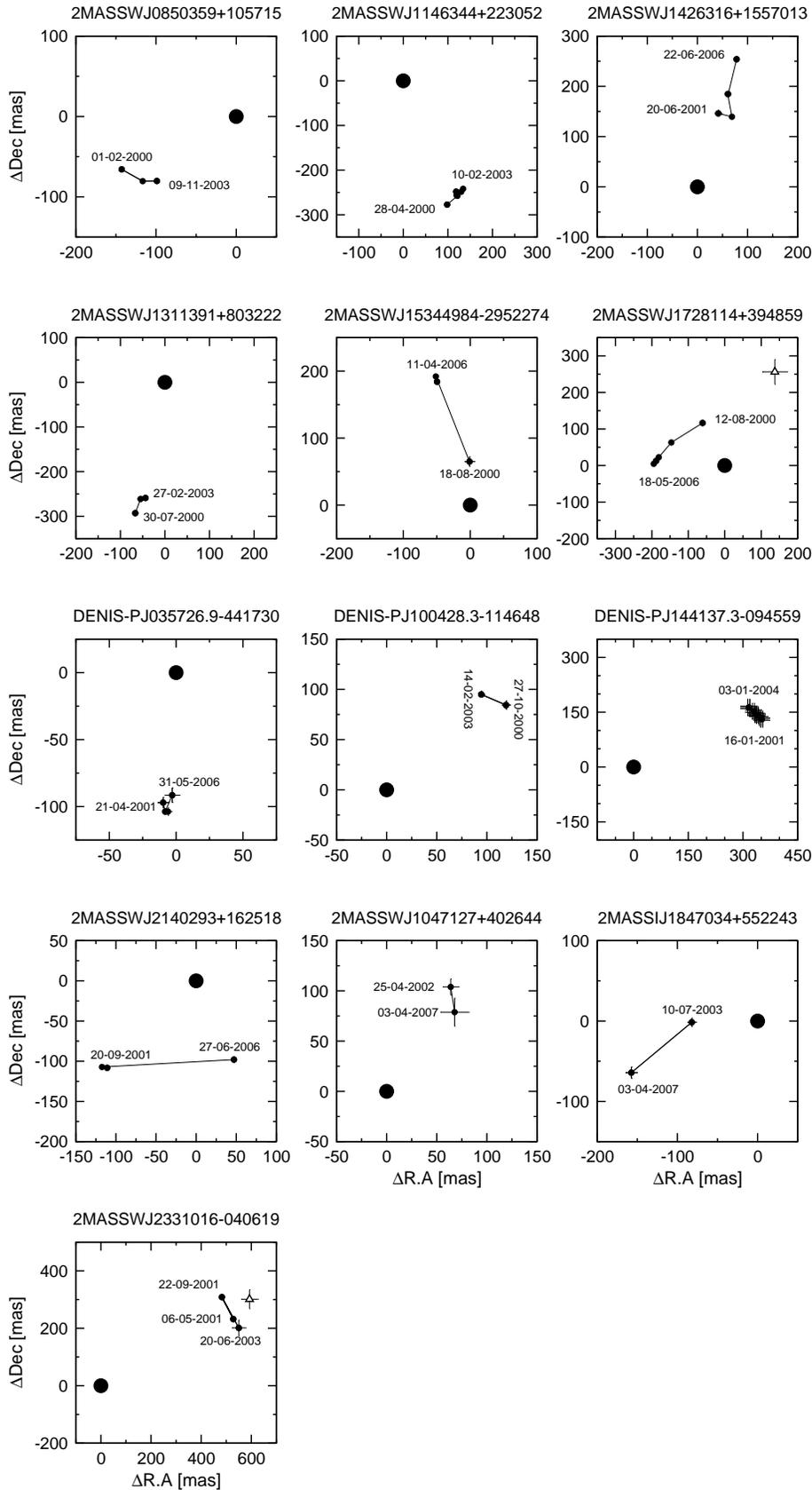}
\caption{Relative orbital motion of the multiple systems presented in this paper. The dates of the first and last epochs are indicated. The primary is represented with a large circle. Esimated uncertainties on the measurements are smaller than the symbols, unless specificied. The instrumental uncertainties are sometimes clearly dominating, as in the case of 2MASSW~J1426316+1557013 (the 2$^{\rm nd}$ epoch deviates significantly, probably because of a large uncertainty in the P.A of the camera on the sky). The open triangle in 2MASSW~J2331016-040619 and 2MASSW~J1728114+394859 panels represents the position that an hypotetic unrelated background star would have had at the last epoch. \label{orbits}}
\end{center}
\end{figure*}

\noindent \emph{2MASSW~J2331016-040619  --} Table \ref{results} and Fig. \ref{orbits} show that the consecutive measurements obtained for this multiple system do not follow a keplerian motion. The Gemini Hokupa'a measurement taken in 2001 indicates a clockwise motion, while the VLT NACO 2003 measurement suggest a counter-clockwise motion. The NACO images, with an exposure time of only 2$\times$30~s during very poor ambient conditions, were of low quality compared to the Gemini and HST images. The faint companion ($\Delta$K=2.44~mag), is barely detected in the NACO images, and the corresponding astrometric measurement is therefore not reliable. We report it for completeness, but it should be considered with caution. 

\noindent \emph{DENIS-P~J035726.9-441730 --} Table \ref{results} and Fig. \ref{orbits} show that the separation was increasing until 2003, and decreasing in the following years. If confirmed (the uncertainties are relatively large), this would mean that the observations were obtained close to the maximum elongation. Assuming a nearly edge-on orbit, as suggested by the current measurements (see Fig. \ref{orbits}), a photometric distance corrected for multiplicity of 22.2~pc \citep{2003AJ....126.1526B}, and a total mass of 0.15~M$_{\sun}$, the semi-major axis of $\approx$104~mas corresponds to a period of $\approx$9~yr. If the distance is larger, i.e., 30~pc, the period remains relatively short ($\approx$15~yr). Hence, DENIS-P~J035726.9-441730 is a promising target for dynamical mass measurement within the next few years.

\subsubsection{Period estimates}

Table \ref{periods} gives estimates of the orbital periods using three different methods. The first column gives the period calculated using Kepler's third law, a statistical scaling of the measured angular separation by 1.26 \citep{1992ApJ...396..178F}, trigonometric, photometric or spectra-photometric derived distances, and mass estimates derived from the spectral types using the spectral type vs T$_{\rm eff}$ of \citet{2002AJ....124.1170D} and the evolutionnary models of \citet{1998A&A...337..403B} for an age of 1~Gyr. The second column gives periods calculated from the fractional change in P.A assuming a circular face-on orbit. The third column gives the period calculated using the ratio of the total motion over the monitored timespan to the quantity 4$\times$maximum projected separation, assuming an edge-on circular orbit (an edge-on circular orbit would traverse the maximum separation approximately four times). Although the hypothesis are strong and numerous and the uncertainties large, the agreement between the different estimates can sometimes provide an idea of the real value of the period, as well as some idea of the inclination or eccentricity of the system. A large difference between the second and third column can indeed imply a large inclination and/or eccentricity. The cases of 2MASSW~J15344984-2952274 and 2MASSW~J1426316+1557013 illustrate the effect of inclination. For these two pairs, the second column (face-on case) gives a period estimate respectively $\approx$12 and 50 times larger than the third (edge-on case) and first columns. A comparison with Fig.~\ref{orbits} shows that these systems are seen nearly edge-on, explaning the strong discrepancy. In some cases, the discrepency between the first column and the two other gives some clue on the eccentricity. In the case of DENIS-P~J035726.9-441730, for which Fig.~\ref{orbits} shows that it was observed at the apastron passage, the large discrepancy between the first column and the other two indicates that the system most probably has an eccentric orbit. The effect of inclination and eccentricity being degenerated, and the uncertainties on the distances and masses being large, it is not possible to go beyond such qualitative discussions.

\section{Future Prospects and conclusions \label{future}}
We present astrometric and photometric results of follow-up observations of 14 UCD binaries. Only half of them are rotating fast enough to provide accurate dynamical masses within the next 15-20 years. The HST, but also the recently commissioned Laser Guide Stars for Adaptive Optics on 8~m class telescopes should allow to discover and follow more UCD binaries, usually too faint and too red even for the IR-WFS of NACO. Some targets not included in the present sample are already part of other on-going programs, and more follow-up observations are likely to be published in the coming months/years. We are currently closely monitoring three additional targets for which dynamical masses will be derived within one year (Bouy et al., in prep.). Another two \citep[$\epsilon-$Indi~Bab and GJ~1001BC, respectively ][]{2004A&A...413.1029M, 2007IAUS..240..329G} are the targets of additional monitoring programs. The total number of "short" period VLM  multiple systems (short meaning periods allowing dynamical mass measurements within 15--20~yr) roughly adds up to a dozen of objects, which has been extracted from original samples of UCDs made of $\approx$140 objects \citep{2003AJ....126.1526B,2003ApJ...587..407C,2003AJ....125.3302G}, i.e. 
the frequency of short-period resolved binaries is about 8\%. If we consider that about 20 binaries (40 masses) are required in order to start calibrating the mass-luminosity relationship, the current study shows that we would need to observe a total of roughly 140/12$\times$20=250 UCDs at high spatial resolution. This estimate means 
that another survey of about 140 more UCDs is needed to discover enough binaries 
that can yield dynamical masses in the near future for a calibration of the mass-luminosity relationship. Even more dynamical masses will be required to extend the study of UCD physical properties to additional parameters, such as age, gravity, and  metallicity. The study of UCDs would therefore greatly benefit from new high spatial resolution surveys dedicated to searching for new multiple systems, and from complementary monitoring programs targeting the shortest period binaries.

\begin{acknowledgements}
H. Bouy acknowledges the funding from the European Commission's Sixth Framework Program as a Marie Curie Outgoing International Fellow (MOIF-CT-2005-8389). We acknowledge the kind and efficient support of Tricia Royle at STScI, Elinor Gates and Bernie Walp at Lick Observatory, and Lowell Tacconi-Garman at ESO and are grateful for their precious help. We also thank our anonymous referee for helping us improving this article. W. Brandner acknowledges support by the Deutsches Zentrum f\"ur Luft- und Raumfahrt (DLR), F\"orderkennzeichen 50 OR 0401. This work is based on observations collected at the European Southern Observatory (Paranal, Chile), programs 70.D-0773, 077.C-0062, 71.C-0327, P.I. Bouy, at the Lick Observatory, with the NASA/ESA Hubble Space Telescope obtained at the Space Telescope Science Institute (STScI), programs GO-9157,GO-9345,GO-9499 and GO-9968, GO-10559 and GO-9451. The STScI is operated by the Association of Universities for Research in Astronomy, Inc., under NASA contract NAS 5-26555. This publication makes use of data products from the Two Micron All Sky Survey, which is a joint project of the University of Massachusetts and the Infrared Processing and Analysis Center/California Institute of Technology, funded by the National Aeronautics and Space Administration and the National Science Foundation. 

\end{acknowledgements}

\begin{center}
\begin{deluxetable}{lcccccc}
\tabletypesize{\small}
\tablecaption{Sample\label{sample}}
\tablewidth{0pt}
\tablehead{
  \colhead{Name}          & SpT     & I   & J   & H   & K  & Ref. }
\startdata
2MASSW~J0850359+105715    & L6      & \nodata     & 16.5 & 15.2 & 14.5 &  \citet{1999ApJ...519..802K} \\
2MASSW~J0920122+351742    & L6.5    & 19.4        & 15.6 & 14.7 & 13.9 & \citet{2000AJ....120..447K}       \\
2MASSW~J1146344+223052    & L2/L2\tablenotemark{a}      & \nodata     & 14.2 & 13.2 & 12.6 &  \citet{1999ApJ...519..802K}  \\
2MASSW~J1426316+1557013   & M8/L1.5\tablenotemark{a}      & 16.5        & 12.9 & 12.2 & 11.7   & \citet{2000AJ....120.1085G}  \\
2MASSW~J1311391+803222    & M7.5/M8\tablenotemark{a}    & 16.2 & 12.8 & 12.1 & 11.7 & \citet{2000AJ....120.1085G}  \\
%2MASSW~J2206228-204705    & L2      & 19.2 & 15.6 & 14.5 & 13.6 \\
2MASSW~J15344984-2952274  & T5.5  & \nodata & 14.9 & 14.9 & 14.8 & \citet{2002ApJ...564..421B}  \\
2MASSW~J1728114+394859    & L7      & \nodata & 16.0 & 14.8 & 13.9 &  \citet{2000AJ....120..447K}  \\ 	
2MASSW~J2331016-040619    & $\approx$L2 & 16.3 & 12.9 & 12.3 & 11.9 &  \citet{2000AJ....120.1085G}    \\
2MASSW~J2140293+162518    & M9 & \nodata & 12.9 & 12.3 & 11.8 &  \citet{2000AJ....120.1085G}      \\
%LHS~2397aB                & M8/L7.5 & 14.6 & 11.9 & 11.2 & 10.7 \\
DENIS-P~J035726.9-441730  & M9/L1.5\tablenotemark{a} & 18.1 & 14.6 & 13.5 & 12.9 &  \citet{1999AJ....118.2466M}   \\
DENIS-P~J100428.3-114648  & M9.5/L0.5\tablenotemark{a} & 18.0 & 14.9 & 14.1 & 13.7 &  \citet{1999AJ....118.2466M} \\
DENIS-P~J144137.3-094559  & L1      & 17.3 & 14.2 & 13.2     & 12.4 &  \citet{1999AJ....118.2466M}  \\
2MASSW~J1047127+402644    & M8  & \nodata & 11.4 & 10.8 & 10.4 &  \citet{2000MNRAS.311..385G}  \\
2MASSI~J1847034+552243    & M7 & \nodata & 1.9  & 11.2 & 10.9 &  \citet{2003AJ....126.2421C}  \\
\enddata
\tablecomments{I,J,K magnitudes of the DENIS objects from the DENIS survey; J,H,K$_{S}$ magnitudes of the 2MASS and LHS objects from the 2MASS survey; I magnitudes for the 2MASS objects from \citet{2003AJ....126.1526B}; H magnitudes for the DENIS objects from the 2MASS survey. If not specified, the spectral type corresponds to that of the unresolved system. Unless specified, the spectral type(s) correspond to those given in the last column reference. }
\tablenotetext{a}{Spectral type from \citet{2006A&A...456..253M}}
%\tablenotetext{b}{Spectral type derived as explained in \citet{2003AJ....126.1526B}}
\end{deluxetable}
\end{center}

\begin{center}
\begin{deluxetable}{lccccc}
\tabletypesize{\small}
\tablecaption{Main characteristics of the instruments used in this study\label{instruments}}
\tablewidth{0pt}
\tablehead{
\colhead{Instrument} & \colhead{Filter} & \colhead{Platescale} & \colhead{Field of view} & \colhead{$\lambda$/D} & \colhead{Resolution} \\
                     &                  & [mas/pixel]          & [\arcsec]               & [mas]                &  [mas]}
\startdata
HST/WFPC2 PC         & F814W     & 45.5                & 44\farcs2$\times$44\farcs2  & 85    & 60  \\
HST/ACS HRC          & F814W     & 25\tablenotemark{a} & 35\farcs4$\times$38\farcs0  & 85 & 40 \\
HST/NICMOS1          & F110M     & 43.2                & 15\farcs7$\times$15\farcs7  & 115   & 90  \\
HST/STIS             & LongPass  & 50.8                & 6\farcs9$\times$6\farcs9   & 75    & 60  \\
VLT/NACO             & Ks        & 13.3                & 13\farcs6$\times$13\farcs6  & 68    & 40  \\
Gemini/Hokupa'a      & Ks        & 20                  & 20\farcs5$\times$20\farcs5  & 68    & 50  \\
Subaru/CIAO          & Ks        & 21.3                & 21\farcs8$\times$21\farcs8  & 68    & 60  \\
Lick/AO              & Ks        & 76                  & 19\farcs4$\times$19\farcs4  & 180   & 110 \\
\enddata
%\tablecomments{}
%\tablenotetext{a}{Estimate of the resolution achieved using PSF fitting on a unity flux ratio multiple system. These values depend on the method used to extract the relative astrometry and photometry and are only indicative.}
\tablenotetext{a}{For pipeline processed data with MultiDrizzle. The "raw" platescale of the ACS/HRC is 28$\times$24.8~mas.}
\end{deluxetable}
\end{center}

%\documentclass{aa}
%\usepackage{deluxetable}
%\begin{document}

%\renewcommand{\arraystretch}{1.5}
\begin{center}
\begin{deluxetable}{lcccccc}
\tabletypesize{\small}
\tablecaption{Relative astrometry and photometry of the mutliple systems\label{results}}
\tablewidth{0pt}
\tablehead{
\colhead{Date of Obs.}  & \colhead{Instrument} & \colhead{Sep. [mas]} &\colhead{P.A [\degr]}  & \colhead{$\Delta$mag} & \colhead{Filter} & \colhead{Ref.\tablenotemark{a}}
}
\startdata
\multicolumn{7}{c}{2MASSW~J0850359+105715}\\
\hline
01-02-2000 & HST/WFPC2 & 157.2$\pm$2.8 & 114.7$\pm$0.3  & 1.47$\pm$0.09 & F814W & (2) \& (3)      \\
21-10-2002 & HST/ACS   & 141.7$\pm$0.9 & 124.6$\pm$0.36 & 1.36$\pm$0.02 & F625W  & (1)      \\
           & HST/ACS   &               &                & 1.21$\pm$0.02 & F775W  & (1)      \\
           & HST/ACS   &               &                & 0.91$\pm$0.08 & F850LP & (1)      \\
09-11-2003 & HST/NICMOS & 127.4$\pm$4.3 & 129.0$\pm$1.8 & 1.10$\pm$0.04 & F110M & (1) \\
\hline
\multicolumn{7}{c}{2MASSW~J0920122+351742}\\
\hline
02-09-2000 & HST/WFPC2 & 75.1$\pm$2.8 & 248.5$\pm$1.2  & 0.88$\pm$0.11 & F814W & (2) \& (3)      \\
19-10-2002 & HST/ACS   & $<$40 & \nodata & \nodata & F625W & (1) \\ 
           & HST/ACS   &       & \nodata & \nodata & F775W & (1) \\
           & HST/ACS   &       & \nodata & \nodata & F850LP & (1) \\
10-03-2003 & HST/STIS  & $<$60 & \nodata & \nodata & LongPass & (8) \\
22-03-2003 & VLT/NACO  & $<$60 & \nodata & \nodata & Ks & (1) \\
03-10-2005 & HST/ACS   & $<$40 & \nodata & \nodata & F814W & (1) \\
08-04-2006 & HST/ACS   & $<$40 & \nodata & \nodata & F814W & (1) \\
\hline
\multicolumn{7}{c}{2MASSW~J1146344+223052}\\
\hline
28-04-2000 & HST/WFPC2 & 294.1$\pm$2.8 & 199.5$\pm$0.3  & 0.75$\pm$0.09 & F814W & (2) \& (3)      \\
08-06-2002 & HST/WFPC2 & 284.8$\pm$2.8 & 205.2$\pm$0.6  & 0.53$\pm$0.09 & F814W & (1)   \\
13-06-2002 & HST/WFPC2 & 282.7$\pm$2.8 & 205.0$\pm$0.6  & 0.55$\pm$0.09 & F814W & (1)  \\
05-05-2003 & HST/WFPC2 & 280.5$\pm$2.8 & 207.6$\pm$0.6  & 0.55$\pm$0.09 & F814W & (1) \\
10-02-2003 & HST/STIS  & 275.1$\pm$2.8 & 205.5$\pm$0.6  & \nodata       & Longpass & (8) \\
13-11-2003 & HST/WFPC2 & 276.5$\pm$2.8 & 209.0$\pm$0.6  & 0.56$\pm$0.09 & F814W & (1) \\
\hline
\multicolumn{7}{c}{2MASSW~J1426316+1557013}\\
\hline
20-06-2001 & Gemini/Hokupa'a & 152$\pm$6     & 344.1$\pm$0.7  & 0.78$\pm$0.05 & J & (4) \\
           & Gemini/Hokupa'a &               &                & 0.70$\pm$0.05 & H & (4) \\
           & Gemini/Hokupa'a &               &                & 0.65$\pm$0.10 & K$_{S}$ & (4) \\
           & Gemini/Hokupa'a &               &                & 0.57$\pm$0.14 & K & (4) \\
19-07-2001 & HST/WFPC2 & 155.6$\pm$1.7 & 333.7$\pm$0.6  & 1.40$\pm$0.09 & F814W & (3) \& (5) \\
           & HST/WFPC2 &               &                & 0.76$\pm$0.11 & F1042M & (3) \& (5) \\
10-03-2003 & HST/ACS   & 194.4$\pm$0.9 & 341.9$\pm$0.3 & 0.99$\pm$0.08 & F625W & (1) \\
           & HST/ACS   &               &                 & 1.22$\pm$0.08 & F775W & (1) \\
           & HST/ACS   &               &                 & 1.31$\pm$0.08 & F850LP & (1) \\
28-04-2003 & HST/STIS  & 194.6$\pm$2.8 & 341.6$\pm$0.8 & \nodata & Longpass & (8) \\
22-06-2006 & VLT/NACO  & 265.8$\pm$1.8 & 342.9$\pm$0.8  & 0.57$\pm$0.02 & Ks & (1) \\
\hline
\multicolumn{7}{c}{2MASSW~J1311391+803222}\\
\hline
30-07-2000 & HST/WFPC2 & 300.4$\pm$3.9 & 167.2$\pm$0.7  & 0.39$\pm$0.07 & F814W & (3) \& (5) \\
           & HST/WFPC2 &               &                & 0.45$\pm$0.09 & F1042M & (3) \& (5) \\
25-04-2002 & Gemini/Hokupa'a & 267$\pm$6 & 168.15$\pm$0.48 & 0.14$\pm$0.05 & K' & (6) \\
27-02-2003 & HST/STIS  & 262.7$\pm$2.8 & 170.4$\pm$0.6  & \nodata       & Longpass & (8)      \\
\hline
\multicolumn{7}{c}{2MASSW~J15344984-2952274}\\
\hline
18-08-2000 & HST/WFPC2 & 65$\pm$7      & 1$\pm$9           & 0.5$\pm$0.3 & F814W & (7) \\
           & HST/WFPC2 &               &                   &             & F1042M & (7) \\
19-01-2006 & HST/ACS   & 198.8$\pm$0.9 & 15.0$\pm$0.1       & 0.26$\pm$0.03 & F814W & (1) \\
11-04-2006 & HST/ACS   & 190.7$\pm$0.9 & 15.1$\pm$0.1      & 0.31$\pm$0.03 & F814W & (1) \\
\hline
\multicolumn{7}{c}{2MASSW~J1728114+394859}\\
\hline
12-08-2000 & HST/WFPC2 & 131.3$\pm$2.8 & 27.6$\pm$1.2      & 0.66$\pm$0.11 & F814W & (3) \& (5) \\
07-09-2003 & HST/NICMOS & 159.6$\pm$4.3 & 66.8$\pm$1.8 & 0.15$\pm$0.04 & F110M & (1) \\
14-08-2005 & HST/ACS    & 182.4$\pm$0.9 & 82.9$\pm$0.3 & 0.45$\pm$0.04 & F814W & (1) \\
18-05-2006 & HST/ACS    & 188.7$\pm$0.9 & 86.2$\pm$0.1 & 0.59$\pm$0.03 & F814W & (1) \\
01-01-2006 & HST/ACS    & 195.0$\pm$0.9 & 88.6$\pm$0.1 & 0.50$\pm$0.03 & F814W & (1) \\
\hline
\multicolumn{7}{c}{DENIS-P~J035726.9-441730}\\
\hline
21-04-2001 & HST/WFPC2 & 97.5$\pm$3.9  & 174.3$\pm$2.3  & 1.23$\pm$0.11 & F675W & (3) \\
           & HST/WFPC2 &               &                & 1.50$\pm$0.11 & F814W & (3) \\
21-08-2002 & HST/ACS   & 103.9$\pm$0.9 & 175.6$\pm$0.5 & 1.09$\pm$0.02 & F625W & (1) \\
           & HST/ACS   &               &                & 1.13$\pm$0.02 & F775W & (1) \\
           & HST/ACS   &               &                & 1.14$\pm$0.02 & F850LP & (1) \\
03-01-2003 & HST/STIS  & 103.9$\pm$2.8 & 176.7$\pm$1.5  & \nodata       & Longpass & (8) \\
13-09-2005 & HST/ACS   & 104.1$\pm$0.9 & 175.5$\pm$0.5 & 1.19$\pm$0.07 & F814W & (1) \\
31-05-2006 & HST/ACS   & 91.5$\pm$5.4  & 178.2$\pm$0.4 & 1.11$\pm$0.04 & F814W & (1) \\
\hline
\multicolumn{7}{c}{DENIS-P~J100428.3-114648}\\
\hline
27-10-2000 & HST/WFPC2 & 146.0$\pm$3.9 & 305.3$\pm$1.5  & 0.25$\pm$0.07 & F675W & (3) \\
           & HST/WFPC2 &               &                & 0.66$\pm$0.11 & F814W & (3) \\
14-02-2003 & HST/STIS  & 133.9$\pm$2.8 & 315.2$\pm$1.2  & \nodata       & Longpass & (8) \\
\hline
\multicolumn{7}{c}{DENIS-P~J144137.3-094559}\\
\hline
16-01-2001 & HST/WFPC2 & 375.3$\pm$2.8 & 290.4$\pm$0.4  & 0.30$\pm$0.07 & F814W & (3) \\
22-05-2001 & HST/WFPC2 & 372.5$\pm$2.8 & 291.3$\pm$0.4  & 0.28$\pm$0.07 & F814W & (3) \\
20-01-2002 & HST/WFPC2 & 367.8$\pm$2.8 & 292.5$\pm$0.4  & 0.26$\pm$0.07 & F814W & (1) \\
29-03-2002 & HST/STIS  & 367.8$\pm$2.8 & 293.0$\pm$0.4  & \nodata       & Longpass & (8) \\
01-05-2002 & HST/WFPC2 & 365.0$\pm$2.8 & 293.2$\pm$0.4  & 0.27$\pm$0.07 & F814W & (1) \\
01-01-2003 & HST/WFPC2 & 362.6$\pm$2.8 & 294.9$\pm$0.4  & 0.27$\pm$0.07 & F814W & (1) \\
03-01-2004 & HST/WFPC2 & 355.6$\pm$2.8 & 297.3$\pm$0.4  & 0.26$\pm$0.07 & F814W & (1) \\
\hline
\multicolumn{7}{c}{2MASSW~J1847034+552243}\\
\hline
10-07-2003 & Subaru/CIAO & 82$\pm$5    & 91.1$\pm$1.4   & 0.16$\pm$0.10 & Ks & (9) \\
03-04-2007 & Lick/AO+LGS & 170$\pm$7   & 112.2$\pm$0.3  & 0.27$\pm$0.15 & Ks & (1) \\
\hline
\multicolumn{7}{c}{2MASSW~J1047127+402644}\\
\hline
25-04-2002 & Gemini/Hokupa'a & 122$\pm$8    & 328.36$\pm$3.75   & 0.50$\pm$0.15 & Ks & (6) \\
           &                 &              &                   & 0.91$\pm$0.20 & H  & (6) \\
03-04-2007 & Lick/AO+LGS         & 106$\pm$14   & 319.3$\pm$1.0  & 0.6$\pm$0.4   & Ks & (1) \\
           &                 &              &                & 1.2$\pm$0.4   & H & (1) \\
\hline
\multicolumn{7}{c}{2MASSW~J2140293+162518}\\
\hline
20-09-2001 & Gemini/Hokupa'a & 155$\pm$5 & 134.30$\pm$0.5 & 0.75$\pm$0.04 & K' & (6) \\
21-05-2001 & HST/WFPC2       & 159.0$\pm$2.8 & 132.4$\pm$0.3 & 1.51$\pm$0.11 & F814W & (3) \& (5) \\ 
           & HST/WFPC2       &               &               & 1.38$\pm$0.11 & F1042M & (3) \& (5) \\
27-06-2006 & VLT/NACO        & 108.7$\pm$1.3 & 205.7$\pm$1.6 & 0.73$\pm$0.02 & Ks & (1) \\
%\hline
%\multicolumn{7}{c}{2MASSW~J2206228-204705}\\
%\hline
%13-08-2000 & HST/WFPC2 & 163.3$\pm$3.9 & 57.3$\pm$1.4  & 0.36$\pm$0.07 & F814W & (3) \\
%           & HST/WFPC2 &               &               & 0.30$\pm$0.07 & F1042M & (3) \\
%22-09-2001 & Gemini/Hokupa'a & 168$\pm$7 & 68.2$\pm$0.5 & 0.08$\pm$0.03 & K     & (6) \\
%20-06-2003 & VLT/NACO  & 157.3$\pm$4.3 & 88.9$\pm$1.6  & 0.00$\pm$0.17 & Ks     & (1) \\
%11-07-2003 & VLT/NACO  & 159.6$\pm$2.9 & 89.5$\pm$1.0  & 0.03$\pm$0.15 & Ks     & (1) \\
%           & VLT/NACO  &               &               & 0.05$\pm$0.15 & H     & (1) \\
%21-08-2003 & VLT/NACO  & 155.2$\pm$4.3 & 93.3$\pm$1.6  & 0.04$\pm$0.17 & Ks     & (1) \\
%28-08-2003 & VLT/NACO  & 152.8$\pm$4.3 & 92.1$\pm$1.6  & 0.02$\pm$0.17 & Ks     & (1) \\
%20-05-2006 & VLT/NACO  & 130.7$\pm$2.6 & 129.6$\pm$1.6 & 0.07$\pm$0.15 & Ks     & (1) \\
%28-06-2006 & VLT/NACO  & 131.5$\pm$2.6 & 131.1$\pm$1.6 & 0.09$\pm$0.15 & Ks     & (1) \\
%29-06-2006 & VLT/NACO  & 135.1$\pm$2.6 & 131.0$\pm$1.6 & 0.19$\pm$0.17 & Ks     & (1) \\
\hline
\multicolumn{7}{c}{2MASSW~J2331016-040619}\\
\hline
06-05-2001 & HST/WFPC2 & 577$\pm$2.8   & 293.7$\pm$0.4  & 3.90$\pm$0.17 & F814W & (3) \\
           & HST/WFPC2 &               &                & 3.54$\pm$0.17 & F1042M & (3)\\
22-09-2001 & Gemini/Hokupa'a & 573$\pm$8 & 302.6$\pm$0.4 &  2.44$\pm$0.03 & K' & (6) \\
20-06-2003 & VLT/NACO  & 586.0$\pm$30 & 290$\pm$3  & \nodata       & Ks     & (1) \\
%\hline
%\multicolumn{7}{c}{LHS~2397aB}\\
%\hline
%04-12-1997 & HST/WFPC2 & 255$\pm$7 & 90.3$\pm$0.8  & 4.34$\pm$0.14 & F814W & (6) \& (7) \\
%07-02-2002 & Gemini/Hokupa'a & 207$\pm$7 & 151.9$\pm$1.2 & 3.15$\pm$0.30 & H & (6)\\
%           &                 &           &               & 2.77$\pm$0.10 & Ks & (6)\\
%26-03-2002 & VLT/NACO  & 164$\pm$4 & 157.3$\pm$2.0 & 3.03$\pm$0.13 & H     & (7) \\
%           &           &           &               & 2.53$\pm$0.03 & Ks    & (7)\\
%31-05-2003 & VLT/NACO  & 187.5$\pm$4 & 187.5$\pm$1.8 & 1.75$\pm$0.05 & L' & (1) \\
%12-01-2006 & VLT/NACO  & 134.8$\pm$1.3 & 276.8$\pm$1.8 & 2.97$\pm$0.02 & Ks & (1) \\
\enddata
\tablenotetext{a}{Reference for the measurement: (1) This work; (2) \citet{2000AJ....119..369R}; (3) \citet{2003AJ....126.1526B}; (4) \citet{2002ApJ...567L..53C}; (5) \citet{2003AJ....125.3302G}; (6) \citet{2003ApJ...587..407C}; (7) \citet{2003ApJ...586..512B}; (8) \citet{2006A&A...456..253M}; (9) \citet{2005ApJ...621.1023S} }
\tablecomments{When several filters are available at the same epoch, the given separations and positions angle correspond to the average of the values measured in the different filters, and the uncertainties to the propagated uncertainties.}
\end{deluxetable}
\end{center}

%\end{document}

\begin{center}
\begin{deluxetable}{lccc}
\tabletypesize{\small}
\tablecaption{Comparison of observed and proper motions \label{motion}}
\tablewidth{0pt}
\tablehead{
  \colhead{Object}        &  \colhead{Proper motion [mas/yr]}  & \colhead{Observed motion B/A [mas/yr]} & \colhead{Ref.}
}
\startdata
2MASSW~J0850359+105715    & 144.7$\pm$2.0                    & 12$\pm$4   & (1) \\
2MASSW~J1047127+402644    & 291$\pm$4                        & 5$\pm$4    & (2)        \\
2MASSW~J1146344+223052    & 96.0$\pm$0.5                     & 14$\pm$4   & (1)  \\
2MASSW~J1426316+1557013   & 97$\pm$2                         & 23$\pm$2   & (5) \\ 
2MASSW~J1311391+803222    & 291$\pm$5                        & 16$\pm$5   & (5)\\ 
DENIS-P~J144137.3-094559  & 204$\pm$18                       & 16$\pm$4   & (4)  \\ 
2MASSW~J15344984-2952274  & 268.8$\pm$1.9                    & 23$\pm$3   & (2)  \\
%2MASSW~J2206228-204705    & 40$\pm$3                        & 31    \\ 
2MASSW~J1728114+394859    & 45.0$\pm$6.4    & 33$\pm$2\tablenotemark{a}   & (3)   \\
2MASSI~J1847034+552243    & 148$\pm$6                        & 26$\pm$3   & (5) \\
2MASSW~J2331016-040619    & 249$\pm$1       & 235$\pm$33\tablenotemark{a}   & (5) \\ 
\hline
2MASSW~J0920122+351742    & \nodata                          & $<$6         & \\
DENIS-P~J035726.9-441730  & \nodata                          & 3$\pm$2    &  \\ 
DENIS-P~J100428.3-114648  & \nodata                          & 12$\pm$6   &  \\
2MASSW~J2140293+162518    & \nodata                          & 33$\pm$2    & \\
\enddata
\tablenotetext{a}{The amplitude of the proper motion and the observed motion are comparable, but the orientations are inconsistent. See also Fig. \ref{orbits}}
\tablecomments{Proper motions from (1) \citet{2002AJ....124.1170D}; (2) \citet{2003AJ....126..975T}; (3) \citet{2004AJ....127.2948V}; (4) \citet{2007arXiv0710.4786J}; (5) USNO-B.1 catalog; Observed motions evaluated using Table \ref{results}, using the most distant measurements together with the corresponding epochs, and assuming a linear motion.}
\end{deluxetable}
\end{center}

\begin{center}
\begin{deluxetable}{lccc}
\tabletypesize{\small}
\tablecaption{Period estimates (in years) using different methods \label{periods}}
\tablewidth{0pt}
\tablehead{
  \colhead{Object}        &  \colhead{Kepler's Law}  & \colhead{Change in P.A} & \colhead{Change in sep.}\\
  \colhead{}        &  \colhead{(at max. elong.)}      & \colhead{(circular face-on)} & \colhead{(circular edge-on)}

}
\startdata 
 2MASSW~J0850359+105715    &         38 &          95 &          80 \\
 2MASSW~J1047127+402644    &         11 &         195 &         151 \\
 2MASSW~J1146344+223052    &         68 &         134 &         237 \\
 2MASSW~J1426316+1557013   &         44 &        1501 &          27 \\
 2MASSW~J1311391+803222    &         61 &         290 &          82 \\
 DENIS-P~J144137.3-094559  &        120 &         155 &         226 \\
 2MASSW~J15344984-2952274  &         16 &         144 &          12 \\
 2MASSW~J1728114+394859    &         21 &          32 &          44 \\
 2MASSI~J1847034+552243    &         18 &          64 &          14 \\
 2MASSW~J2331016-040619    &        147 &          15 &         220 \\
 2MASSW~J0920122+351742    &          5 &      \nodata & \nodata \\
 DENIS-P~J035726.9-441730  &          8 &        1319 &         260 \\
 DENIS-P~J100428.3-114648  &         44 &          84 &         111 \\
 2MASSW~J2140293+162518    &         19 &          24 &          64 \\
\enddata
%\tablecomments{}
\end{deluxetable}
\end{center}

\end{document}